# Area Formation and Content Assignment for LTE Broadcasting

Claudio Casetti, Carla-Fabiana Chiasserini, Francesco Malandrino, Carlo Borgiattino

**Abstract**

Broadcasting and multicasting services in LTE networks are shaping up to be an effective way to provide popular content. A key requirement is that cells are aggregated into areas where a tight time synchronization among transmissions is enforced, so as to broadcast the same radio resources. Our paper addresses a facet of LTE broadcasting that has so far received little attention: the creation of broadcasting areas and the assignment of content to them in order to efficiently exploit radio resources and satisfy user requests. Our original clustering approach, named Single-Content Fusion, achieves these goals by initially aggregating cells into single-content areas and maximizing cell similarity in content interests. Aggregated areas are then merged into multiple-content areas by virtue of similar spatial coverage. We show the validity of our solution pointing out the advantages it provides in comparison to other approaches. We also discuss the impact of various system factors (e.g., number of served users, broadcast data rate, area size) and the scalability of our proposal in large, realistic scenarios with both static and time-varying user interest.

*Francesco Malandrino is the corresponding author for this paper; email: francesco.malandrino@polito.it

## 1. Introduction

Cellular network operators all over the world are partnering with content providers, such as television stations, or content owners, such a sport leagues and franchises, to offer a wide-ranging array of multimedia streams to their subscribers. While the traditional way of delivering such a service on cellular networks was through multiple unicast flows, native LTE broadcasting and multicasting solutions are finally being rolled out. No longer hampered by a needless waste of radio resources and an excessive strain on the network, broadcasting services, traditionally limited to radio and television, are thus revamped and added increased reach, scalability and flexibility by LTE broadcast. Often cited commercial use cases [1] include: breaking news, live streaming of popular sport events, information feeds to standalone displays in crowded areas. More creative, innovative applications can be listed: augmented experience and in-event advertising on user mobiles at pop concerts and other live events; warnings, alerts and safe evacuation instructions in disaster scenarios; departure/arrivals information at airports and railway stations; periodic software and firmware push updates for unattended devices.

LTE broadcast is especially appealing because users do not have to equip themselves with additional hardware: no special chipset is required and even low-end devices can receive broadcast content if their operators provide it. The operators too have reasons to be attracted by LTE broadcast, since the existing LTE infrastructure can be fully reused, with high-SNR service achieved by a user-level combination of signals from different eNBs. It goes without saying that appealing business models can be envisioned when the simultaneous reaching of thousands of users with limited radio resources is made possible.

3GPP standardization of broadcasting services, or Multimedia Broadcast Multicast Service (eMBMS) [2, 3], began with Release 6 of the standard and reached full maturity with Release 9. It has since been a fixed staple of each successive release. The eMBMS standard defines the procedures involved in the simultaneous transmission of identical content by nearby cells, composing an area called MBSFN (Multimedia Broadcast Single Frequency Network). All cells in an MBSFN use the same radio resources for broadcasting services (hence the Single Frequency in the acronym) in order to exploit constructive interference between signals coming from different sources. As shown in Fig. 1, MBSFNs can overlap in space and are part of a larger *synchronization area*, whose eNBs are required to be synchronized in time. MBSFN area formation decisions within each synchronization area are made centrally at an entity called *area controller*, which is also in charge of collecting information on the users' interests. In principle, a synchronization area can be as small as a few cells, or as wide as an entire country [1]. In each cell of an MBSFN, broadcast and unicast channels coexist, sharing the cell capacity. Broadcasting-related communication between users, eNBs and area controllers takes place through standardized LTE control messages; the associated overhead is negligible if compared to the benefit of MBSFN broadcasting.

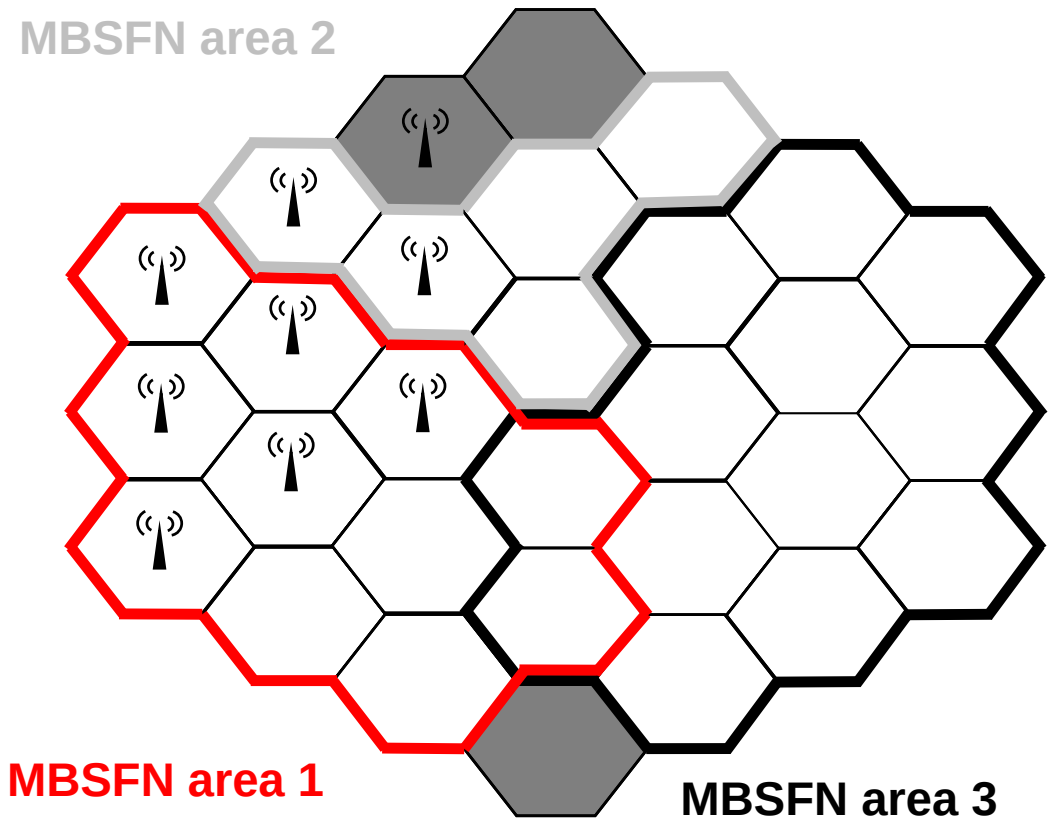

Figure 1: Example of a synchronization area including three MBSFNs (outlined in red, light gray and black). Dark Gray cells are not included in any MBSFN and deliver content via unicast transfers only.

In spite of the attractiveness of LTE broadcasting for all involved actors, lingering challenges still exist, among which physical-layer issues [4] and resource allocation [5, 6, 7, 8] have been widely investigated in the literature. A fundamental aspect that so far has been scarcely addressed is instead MBSFN area formation. In this work, we fill this gap by answering the following questions: (i) How should cells be grouped into MBSFNs? (ii) What kind of content should they convey? (iii) What user demand for broadcast content should be served through eMBMS and what via unicast transmissions?

As shown in [9], finding an optimal solution to the above issues implies solving a mixed integer linear programming (MILP) formulation, whose complexity prevents any application to real-world scenarios. Instead, we adopt a novel content-aware clustering approach, based on the following key observation. In order to optimally form MBSFNs, it is crucial that cells, whose users are interested in the same set of content items, are grouped together. To this end, a metric should be defined that captures the *similarity of user interests* between any pair of cells. Cells can then be clustered with the aim to maximize such a metric.

Similarity metrics normally account for multiple content items [13, 4, 6]. Contrary to most existing works, in this paper such a similarity is computed on a per-content item basis. Our rationale is that evaluating a multi-dimensional similarity metric that accounts for multiple items at the same time increases the complexity and, ultimately, waters down any large difference in content interest between cells. Instead, computing such a metric for one content item at a time, the similarity between cells can be gauged much more accurately. This intuition leads to a two-step procedure, named Single-Content Fusion (SCF), which initially creates single-content areas and, then, merges areas that significantly overlap in space. As a result, the procedure returns multi-content MBSFNs that include cells with very similar content interests and broadcast the most demanded items thus maximizing the system throughput.

We compare our SCF approach to an algorithm, referred to as Multi-Content Fusion (MCF), that leverages a multi-dimensional distance metric computed over multiple content items. An extensive simulation analysis confirms our intuition, showing the superiority of SCF over MCF.

The organization of the paper is as follows. Sec. 2 discusses previous work and highlights the novelty of our contribution. Sec. 3 introduces the system model and the constraints imposed by 3GPP on the MBSFN configuration. Sec. 4 describes the proposed SCF algorithm and discusses its complexity, while Sec. 5 presents the MCF scheme that is used for comparison. Numerical results showing the benefits of our approach in both static and dynamic scenarios and giving useful insights on how to successfully support eMBMS are presented in Sec. 6. Finally, Sec. 7 concludes the paper and draws directions for future research.

## 2. Previous work and contribution

Multimedia Broadcast Multicast Service in cellular networks was initially introduced for 3G systems in Rel. 6 of the standard, while eMBMS has been defined for LTE systems starting with Rel. 9, and enhancements have appeared in all subsequent releases [2, 3]. In addition to 3GPP standards, eMBMS has attracted plenty of attention in the literature.

Of particular relevance to our work are [5] and [6], which aim to maximize the throughput of broadcast/multicast services in a single-cell scenario. The former sacrifices users with low data rate while guaranteeing the desired service coverage, whereas the latter optimally forms multicast groups based on the users data rate. In [12], several radio resource management policies for MBMS data delivery have been evaluated in terms of aggregate data rate, fairness and spectral efficiency. However, also this work is limited to an MBSFN area composed of a single cell. An adaptive modulation and coding, as well as frequency scheduling, scheme for video transmission with scalable video coding is presented in [7]. A framework that maximizes the revenue of mobile operators is proposed in [8], where broadcast or unicast transmissions are scheduled for each content based on its characteristics.

To our knowledge, the only previous works dealing with aspects related to the formation or the configuration of MBSFNs are [13, 15, 9, 10, 11]. The analysis in [13] assumes eNBs to be distributed according to a random point process and studies the impact of the minimum spatial separation between eNBs on the network outage probability. As for [15], it has the merit of having first dealt with the problem of area formation, however it sketches a genetic algorithm that makes several major simplifications. Firstly, the algorithm in [15] does not account for the topological and technical constraints imposed by the eMBMS standard, including the requirement that MBSNF areas be contiguous. Secondly, it is limited to area formation and it does not determine the content items to be broadcasted in each area. Thirdly, it maximizes the system coverage in terms of number of served users, but it does not deal with system throughput. Similarly, [9] casts an optimization problem that maximizes the system coverage as defined above. That work shows that, even in the simple case where throughput is neglected, forming optimal MBSFNs is a MILP problem. A similar formulation is obtained by the authors of [10], who also propose and evaluate two alternative solution strategies, based on an iterative approach (fuse-and-refine) and genetic algorithms respectively. Finally, the more recent work [11] targets the area formation problem by assigning a single frequency for each area, and enlisting the help of user devices to assist the broadcast via device-to-device transfers.

*Our contribution.* Unlike previous work, we devise an algorithm that *jointly* forms MBSFN areas and determines which content should be broadcast by which MBSFN. Areas and content to be broadcasted are determined so as to (i) meet the system constraints imposed by the 3GPP specifications, (ii) maximize the overall system throughput when both unicast and broadcast traffic are accounted for, and (iii) satisfy a target fraction (in the following set to 1) of broadcast demand. We reach these goals in a highly-efficient fashion, capable of handling large-scale scenarios. It is worth mentioning that a conference version of this work has appeared in [14], where however only a static demand and user interest scenario was considered.

## 3. eMBMS network model

In this section, we describe our mathematical model of eMBMS networks. Specifically, Sec. 3.1 presents our system model, i.e., how the entities involved in eMBMS networks are described in terms of sets, elements thereof, and properties of said elements. Then, Sec. 3.2, recalls the constraints to which area formation decisions are subject (as stated in 3GPP standards) and show how they can be expressed in our system model. Finally, in Sec. 3.3 and Sec. 3.4 we use our system model to represent the effect of the area formation decisions on the system performance.

### *3.1. System model*

We focus on a synchronization area covering a set of cells, denoted by $\mathcal{C}$. We denote by $\mathcal{E}$ the set of all user equipment (UE) in the synchronization area. Through the so-called counting procedure [2], eNBs collect feedback from the UEs on whether they are interested or not in a list of content that eNBs advertise as potential broadcast items. Let $\mathcal{I}$ be the set of such items and $\mathcal{E}_{B,c} \subseteq \mathcal{E}$ the set of users in cell $c$ interested in them, i.e., contributing to broadcast demand. Let $\mathcal{E}_{U,c} \subseteq \mathcal{E}$ be the set of UEs in cell $c$ contributing to unicast demand.

As mentioned, a synchronization area may include a number of MBSFN areas (see Fig. 1). The set of MBSFNs that are activated over the synchronization area is indicated by $\mathcal{M}$. Each MBSFN area, $m \in \mathcal{M}$, includes contiguous cells that broadcast the same set of content items using the same radio resources. Hence, $m \subseteq \mathcal{C}$. Within an MBSFN area, the network may decide (i) to serve all or just a fraction of broadcast items requested by UEs and (ii) all or just a fraction of UEs requesting a broadcasted item. In the latter case, the leftover broadcast demand may be served through unicast service.

Downlink LTE radio resources are grouped into resource blocks (RBs), each including 12 consecutive subcarriers and lasting for 0.5 ms. In practice, however, the periodicity with which resource scheduling is performed is equal to one subframe, i.e., 1 ms. The time period corresponding to 10 subframes is called frame. We assume the broadcast traffic periodicity to be equal to 1 frame and, for simplicity, we *take one frame as reference period.* Let $R$ be the total number of RBs that each cell can use per frame (thus, $R/10$ is the number of radio resources per subframe).

RB allocation is determined by both channel quality and service data rate. The channel quality depends on signal propagation conditions as well as on the interference received from neighboring eNBs. The channel quality experienced by a UE is fed back to the network through the channel quality indicator (CQI). This value determines the Modulation and Coding Scheme (MCS) used for traffic delivery to the UE. In case of broadcast service, it is typically the UE that experiences the worst CQI that drives the MCS selection for the broadcast transmission. The MCS, in its turn, determines the number of bits carried in each RB. It follows that, given the content service rate (e.g., a video stream requiring a 1-Mb/s bandwidth), the number of RBs that have to be allocated to support such a service depends on the CQI experienced by the recipient UEs. Thus, a broadcast content service is feasible if enough RBs are available to match its service rate, given the CQI of the worst-channel UE.

With reference to *broadcast service* in MBSFN $m$, we will denote by $B_m^i$ the average number of RBs per frame required by the delivery of item $i$, and by $b_m^i(k)$ the number of required RBs in subframe $k$. $B_m^i$, hence $b_m^i(k)$, is equal to 0 if item $i$ is not broadcasted in area $m$. If it is broadcasted, its value depends on the CQI of the users that are selected to be served through eMBMS, as explained above. As for the delivery of *unicast traffic* to a UE $j \in \mathcal{E}_{U,c}$ in cell $c \in \mathcal{C}$, $U_{j,c}$ will denote the average number of allocated RBs per frame and $u_{j,c}(k)$ the number of RBs per subframe $k$. Finally, in case broadcast content is delivered to a UE through unicast service, let $X_{j,c}^i$ be the average number of RBs required for unicast delivery of broadcast content $i$ toward user $j$ in cell $c$, and $x_{j,c}^i(k)$ the number of required RBs in subframe $k$. If UE $j$ is not served through unicast, the $U_{j,c}$, $u_{j,c}(k)$, $X_{j,c}^i$ and $x_{j,c}^i(k)$ quantities are equal to 0.

*3.2. System constraints*

According to 3GPP [2, 3], there are some system constraints guiding the formation of the set $\mathcal{M}$ of MBSFNs and content broadcasting therein. They can be expressed as follows.

*(i)* Each MBSFN is denoted by an identifier, whose value ranges between 0 and Max_MBSFN $-$ 1, with Max_MBSFN $=$ 256 [2, 16]. According to the standard [2], such an id might be not unique within a synchronization area: the only clear constraint is that two neighboring areas cannot be assigned the same id. Neighboring MBSFNs are areas that are overlapping or adjacent. More formally, by denoting with $\mathcal{N}_i$ the set of neighboring areas of MBSFN $m_i$ and by $|\mathcal{N}_i|$ the set cardinality, we have: $|\mathcal{N}_i| \leq$ Max_MBSFN, $\forall m_i \in \mathcal{M}$. Other references, however, consider a more stringent constraint and impose that the number of MBSFNs that can be formed in a synchronization area cannot exceed Max_MBSFN, i.e., $|\mathcal{M}| \leq$ Max_MBSFN. In the following, we consider the first interpretation, as it is more conservative and certainly compliant with the current standard. Nevertheless, since the issue has not been fully clarified yet, we also investigate the impact of both the above constraints on the system performance.

*(ii)* A cell can belong to at most 8 MBSFNs, i.e.,

$$\sum_{m \in \mathcal{M}} \mathbb{1}_{c,m} \leq 8 \quad \forall c \in \mathcal{C}\,,$$

where the indicator function $\mathbb{1}_{c,m}$ is equal to 1 if cell $c$ belongs to area $m$ and 0 otherwise.

*(iii)* In each frame and in each cell $c \in \mathcal{C}$, at most 60% of the available RBs can be allocated to broadcast traffic, i.e.,

$$\sum_{i \in \mathcal{I}} \sum_{m \in \mathcal{M}} B_m^i \mathbb{1}_{c,m} \leq 0.6R \quad \forall c \in \mathcal{C}\,.$$

*(iv)* In each subframe $k$ and in each cell $c$, unicast and broadcast traffic cannot use more than the total number of available resources $(R/10)$, i.e.,

$$\sum_{i\in\mathcal{I}}\left(\sum_{m\in\mathcal{M}} b^i_m(k)\mathbb{1}_{c,m} + \sum_{j\in\mathcal{E}_{B,c}} x^i_{j,c}(k)\right) + \sum_{j\in\mathcal{E}_{U,c}} u_{j,c}(k) \le R/10 \quad \forall c\in\mathcal{C},\ \forall k\,.$$

*3.3. Throughput of a synchronization area in a static scenario*

Our goal is to devise a scheme for MBSFN formation and content-to-area assignment that meets the above constraints and maximizes the overall throughput of the synchronization area. The throughput computation requires that a formalization of the different components, i.e., the throughput of broadcast and unicast traffic.

Given the set of MBSFN areas, $\mathcal{M}$, the aggregate broadcast throughput per frame over all areas in $\mathcal{M}$ is given by:

$$T_B^{(B)} = \sum_{c\in\mathcal{C}}\sum_{i\in\mathcal{I}}\sum_{m\in\mathcal{M}} B^i_m w^i_c \rho^i_m \mathbb{1}_{c,m}\,, \tag{1}$$

where $w^i_c$ is the number of UEs in cell $c$ that receive item $i$ via eMBMS. The above expression represents the total broadcast throughput at the receivers as it accounts for the fact that, in each cell, a given content item $i$ may be received by multiple users interested in that content. The quantity $\rho^i_m$ is the per-RB throughput that is obtained for content item $i$ in MBSFN $m$. Such value clearly depends on the data rate at which the broadcast transmission is performed.

As mentioned, any UE, $j\in\mathcal{E}_{B,c}$, that is interested in broadcast content but is left out of eMBMS should be served through unicast. The throughput corresponding to data transfers of this type is given by:

$$T_B^{(U)} = \sum_{c\in\mathcal{C}}\sum_{i\in\mathcal{I}}\sum_{j\in\mathcal{E}_{B,c}} X^i_{j,c}\rho_{j,c}\,, \tag{2}$$

where $\rho_{j,c}$ is the per-RB throughput of UE $j$ in cell $c$, which is an input parameter.

The aggregate unicast throughput over the synchronization area can be computed based on the number of RBs in a frame that are allocated to unicast demand and on the user throughput per RB ($\rho_{j,c}$). Again, the latter depends on the CQI of unicast users. We can write:

$$T_U = \sum_{c \in \mathcal{C}} \sum_{j \in \mathcal{E}_{U,c}} U_{j,c} \rho_{j,c} \, . \tag{3}$$

The overall throughput of the synchronization area is:

$$T = T_B^{(B)} + T_B^{(U)} + T_U \, . \tag{4}$$

Recall that our aim is to maximize (4) while meeting constraints *(i)–(iv)* and targeting to serve 100% of broadcast demand. Although the problem we pose differs from the ones so far addressed in the literature [15, 9], solving it to the optimum would involve a MILP formulation as in [9], with a complexity that hinders its application to real-world scenarios. Thus, in order to achieve our goal, we introduce the SCF procedure, which accomplishes the following tasks with low complexity:

(1) determining the MBSFN areas (hence the values of $\mathbb{1}_{c,m}$);

(2) assigning content to areas (i.e, determining whether $B_m^i = 0$ or not);

(3) selecting which users have to receive the generic content item $i$ through broadcasting ($w_c^i$) and which have to receive it through unicast; this decision determines the values of $X_{j,c}$ as well as the values taken by $B_m^i$ and $\rho_m^i$.

With reference to the first point, we stress that large areas (i.e., MBSFNs including many cells) can leverage constructive interference at broadcast users located in inner cells of the area. Indeed, the transmissions from neighboring eNBs that belong to the same MBSFN use the same radio resources to broadcast the same content, hence boosting the CQI. On the other hand, in presence of even one single low-rate user, the larger the MBSFN, the higher the number of cells affected by this inefficiency. As for the third point, eMBMS is convenient when the number of users interested in content $i$ is sufficiently high and their channel quality is good enough (so that efficient MCS schemes can be used). In all other cases, whether to serve a content via broadcast or unicast has to be carefully evaluated.

The allocation of unicast resources ($U_{j,c}$) is not handled by the SCF procedure since our focus is on broadcast demand. We will simply assume that any RB left after the allocation of broadcast service is allocated through the standard Proportional Fair scheduling algorithm.

### *3.4. Throughput of a synchronization area in a dynamic scenario*

So far, we have just considered a static scenario where decisions are taken only once. In the real world, users move and their interests change over time, and content may become unavailable or lose the former appeal. The activated MBSFN areas may thus be unfit to address demand changes, and new decisions should be made. Let $P$ be the period of time during which area formation decisions are valid, and after which areas are redrawn. Assuming our model and algorithms are run at $t_0$ for the first time, we need to capture some additional conditions, namely:

- some content items may cease to exist between $t_0$ and $t_0 + P$;
- some content items may start between $t_0$ and $t_0 + P$;
- the number of users interested in a content item may change between $t_0$ and $t_0 + P$.

The first two conditions mostly concern streaming events whose duration is roughly known in advance, e.g., football games. The fraction of the time interval $[t_0, t_0 + P]$ for which item $i \in \mathcal{I}$ exists can be modeled in advance as a coefficient $\epsilon^i \in [0, 1]$. Then, we can re-write (1) as follows:

$$T_B^{(B)} = \sum_{c\in\mathcal{C}} \sum_{i\in\mathcal{I}} \sum_{m\in\mathcal{M}} B_m^i w_c^i \rho_m^i \epsilon^i \mathbb{1}_{c,m}.$$

The last condition means that the number of UEs in cell $c$ that receive content $i$ via eMBMS, $w_c^i$, is no longer constant between $t_0$ and $t_0+P$, and shall be replaced by the integral from $t_0$ to $t_0+P$ of its time-dependent version $w_c^i(t)$. The broadcast throughput can thus be rewritten as:

$$T_B^{(B)} = \sum_{c\in\mathcal{C}}\sum_{i\in\mathcal{I}}\sum_{m\in\mathcal{M}} B_m^i \rho_m^i \epsilon^i \mathbb{1}_{c,m} \int_{t_0}^{t_0+P} w_c^i(t)\mathrm{d}t\,.$$

Sadly, there is no practical way to know $w_c^i(t)$ with sufficient precision. We are more likely to have *estimates* of the quantity $\int_{t_0}^{t_0+P} w_c^i(t)\mathrm{d}t$, e.g., the expected number of users interested in content $i$ [1]:

$$\tilde{w}_c^i = \mathbb{E}\left[\int_{t_0}^{t_0+P} w_c^i(t)\mathrm{d}t\right].$$

We therefore use such estimates in (1), obtaining:

$$T_B^{(B)} = \sum_{c\in\mathcal{C}}\sum_{i\in\mathcal{I}}\sum_{m\in\mathcal{M}} B_m^i \tilde{w}_c^i \rho_m^i \epsilon^i \mathbb{1}_{c,m}. \tag{5}$$

This new expression for $T_B^{(B)}$ provides a simple, yet effective, way to account for the challenges introduced by a dynamic scenario. Their impact on the system performance is then evaluated in Section 6.2.

## 4. The Single-Content Fusion procedure

In theory, solving the optimization problem presented in the previous section would allow us to make optimal area formation and content selection decisions. However, the complexity of such a problem, which falls in the mixed-integer lin-

[1] If we have some knowledge of the distribution of $w_c^i$, we may choose to base our decisions on some quantile, such as:

$$\tilde{W}_c^i(0.1) = x\colon \mathbb{P}\left[\int_{t_0}^{t_0+P} w_c^i(t)\mathrm{d}t \le x\right] = 0.1.$$

which yields very conservative broadcast decisions, i.e., broadcasting only those content items that are likely to be popular enough.

ear programming (MILP) category, renders such an approach infeasible. Therefore, in this section we present a heuristic procedure called *single-content fusion* (SCF), making the same decisions in a much more efficient – albeit potentially suboptimal – manner.

The SCF procedure consists of four algorithms: Cell Aggregation, Hill Climbing, Rate Increase and Area Fusion. The block diagram in Fig. 2 depicts such main blocks and the sequence according to which they are executed, along with the input/output of each algorithm.

The procedure operates by first aggregating the initial set of cells, $\mathcal{C}$, into several potential single-content MBSFNs, $\mathcal{A}$, based on user interests (Cell Aggregation). These MBSFNs may be spatially overlapping and, while they nominally satisfy the whole content demand, they are likely to be overdimensioned and not compliant with the system constraints. Thus, as a second step (Hill Climbing), the MBSFNs are individually evaluated to verify the adherence to constraints *(ii)–(iv)* introduced in the previous section and to gauge their potential throughput. The goal is to activate only the single-content MBSFN areas that meet the system constraints and maximize the overall throughput (output as $\mathcal{M}^{(1)}$ areas and $T^{(1)}$ throughput, respectively).

Next, the Rate Increase algorithm purges the set of broadcast users of those with low bit rate so as to boost the overall system throughput. In other words, if more convenient, users with poor CQI are left out of the eMBMS and may be served via unicast transfers. A new set of single-content areas $\mathcal{M}^{(2)}$ is thus output, along with the overall throughput ($T^{(2)}$).

Finally, the Area Fusion algorithm creates multi-content areas by merging those that fully overlap. In case more areas than allowed (e.g., 256, as in constraint *(i)*) are activated, the algorithm keeps merging areas, starting from those exhibiting the largest spatial overlap, and yields the final set $\mathcal{M}$. As a result, the SCF procedure forms MBSFNs, determines which content has to be broadcast by which MBSFN and selects the UEs that should receive it, so that the overall system throughput is maximized.

We remark that, as suggested by its name, one of the main steps of the SCF

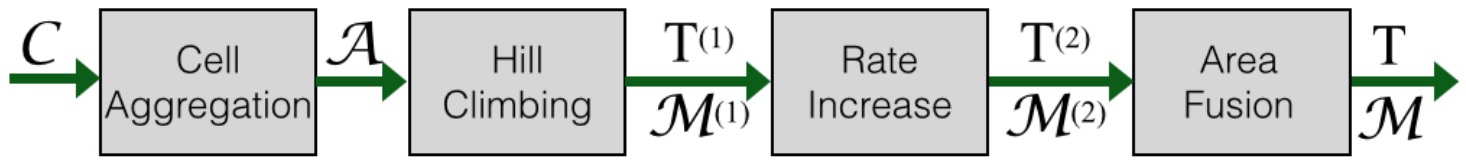

Figure 2: Block diagram of the SCF procedure, including inputs and outputs of each algorithm. $T$ and $\mathcal{M}$ denote the same quantities but their values are changed by each block, hence the (1),(2) apices.

procedure relies on a hill-climbing algorithm. As confirmed by our results in Sec. 6, this approach allows a fast, efficient solution to an otherwise complex problem. The term *hill-climbing* denotes a greedy approach to feature-selection problems such that, if the algorithm execution is interrupted at any point, the partial solution is feasible [17]. Also, for problems with submodular utility, hill-climbing algorithms are proven to provide solutions within $1-\frac{1}{e}\approx 0.63$ from the optimum. The submodularity property requires that the utility (the throughput increase in our case) of selecting a feature (activating an additional area) is non-increasing as the set of already-active features (selected areas) becomes bigger – a property that, as discussed in 4.1, is satisfied in our case.

In the following subsections, we provide a detailed description of the four algorithms and of their complexity.

**Algorithm 1** Cell Aggregation

**Require:** $\mathcal{C}$, $\mathcal{E}_{B,c}$

$\mathcal{A} \leftarrow \emptyset$
**for** $i \in \mathcal{I}$ **do**
  $\widetilde{\mathcal{C}} \leftarrow \emptyset$
  **for** $c \in \mathcal{C}$ **do**
    **if** (no. users in $c$ interested in $i$) $\geq \tau$ **then**
      $\widetilde{\mathcal{C}} \leftarrow \widetilde{\mathcal{C}} \cup \{c\}$
  **while** $\widetilde{\mathcal{C}} \neq \emptyset$ **do**
    Select $c \in \widetilde{\mathcal{C}}$
    $a \leftarrow \{c\}$, $\widetilde{\mathcal{C}} \leftarrow \widetilde{\mathcal{C}} \setminus \{c\}$
    $\mathcal{N} = \mathit{FindNeighbors}(c)$
    **while** $\mathcal{N} \cap \widetilde{\mathcal{C}} \neq \emptyset$ **do**
      $a \leftarrow a \cup \{\mathcal{N} \cap \widetilde{\mathcal{C}}\}$, $\widetilde{\mathcal{C}} \leftarrow \widetilde{\mathcal{C}} \setminus \{\mathcal{N} \cap \widetilde{\mathcal{C}}\}$
      $\mathcal{N} = \mathit{FindNeighbors}(\mathcal{N})$
    $\mathcal{A} \leftarrow \mathcal{A} \cup \{a\}$
**return** $\mathcal{A}$

### *4.1. Algorithm description*

The first algorithm is dubbed *Cell Aggregation* and its pseudocode is listed as Algorithm 1. Its purpose is to define potential cell aggregations that can then be turned into MBSFN areas with similar content demands. First it selects, for every content item $i \in \mathcal{I}$, cells carrying a minimum ($\tau$) number of users interested in content item $i$ (lines 5–6). Then, it identifies aggregations of such selected cells by checking if they are each others' neighbors (lines 10–13). Every time a contiguous set of cells, $a$, is collected and cannot be further expanded, it is added to set $\mathcal{A}$ as a candidate area (line 14). The final set $\mathcal{A}$ of candidate areas is the output of the algorithm. We remark that, though unlikely, $\mathcal{A}$ can also include single-cell areas, provided each satisfies the condition on the minimum number of users interested in content $i$ in line 5.

The set $\mathcal{A}$ of potential MBSFN areas is input to the *Hill Climbing* algorithm (Algorithm 2), whose task is to select in which areas the activation of the broadcasting service is convenient. This task is accomplished in several testing rounds, each adding one more MBSFN area to the activation set $\mathcal{M}$ (i.e., the set of MBSFN areas where broadcast is not only feasible, but also deemed convenient from a system viewpoint).

In the first round, for each area $a \in \mathcal{A}$, the algorithm verifies that the addition of $a$ to $\mathcal{M}$ does not violate constraints *(ii)-(iv)* set forth in Sec. 3.2 (line 4). Constraint *(i)* will be handled later by the Area Fusion algorithm. If those constraints are met, the algorithm computes the overall system throughput (as $T = T_B^{(B)} + T_B^{(U)} + T_U$) resulting from the addition of $a$ to the current set of MBSFNs (line 5).

In order to compute the throughput as in Equations (1)–(4), we set $w_c^i$ so that it includes all users in cell $c$ interested in item $i$. Then, using knowledge of the channel quality experienced by users, the rest of the parameters can be computed. As for the $X_{j,c}^i$ values, we assume that allocation is done sorting by decreasing CQI value the users that have to receive broadcast content via unicast. We remark that the system throughput is independently computed in line 5 for every possible addition of a single area $a$. Also, the throughput

**Algorithm 2** Hill Climbing

**Require:** $\mathcal{A}$, $\mathcal{E}_{B,c}$

1: $\mathcal{M} \leftarrow \emptyset$, $T = 0$
2: **while** $\mathcal{A} \neq \emptyset$ **do**
3:     **for** $a \in \mathcal{A}$ **do**
4:         **if** *VerifyConstraints*$(\mathcal{M} \cup \{a\})$ == True **then**
5:             $\widetilde{T} = \text{ComputeThroughput}(\mathcal{M} \cup \{a\}, \mathcal{E}_{B,c})$
6:             **if** $\widetilde{T} > T$ **then**
7:                 $T = \widetilde{T}$, $a^* \leftarrow a$
8:     **if** $a^*$ is found **then**
9:         $\mathcal{M} \leftarrow \mathcal{M} \cup \{a^*\}, \quad \mathcal{A} \leftarrow \mathcal{A} \setminus \{a^*\}$
10:     **else**
11:         break
12: **return** $T$, $\mathcal{M}$

increase due to the addition of $a$ is independent of $\mathcal{M}$. This independence yields the submodularity property required by Hill Climbing algorithms to provide close-to-the-optimum solutions.

The area $a^*$ is removed from the potential set $\mathcal{A}$ and permanently added to the activation set $\mathcal{M}$ (line 9). After the whole set $\mathcal{A}$ has been tested, a new round is started. If no area can be added, the final set of MBSFN areas is output and the algorithm terminates (line 11).

**Algorithm 3** Rate Increase

**Require:** $\mathcal{M}$, $T$, $\mathcal{E}_{B,c}$, $\mathcal{S} = \{\text{sorted user data rates}\}$,

**for** $s \in \mathcal{S}$ **do**
    $\mathcal{A}_s, \mathcal{E}_s = \text{FindRateUsers}(\mathcal{M}, s, \mathcal{E}_{B,c})$
    $\widetilde{\mathcal{E}}_{B,c} \leftarrow \mathcal{E}_{B,c} \setminus \mathcal{E}_s$
    $\hat{\mathcal{A}} = \text{CellAggregation}(\mathcal{A}_s, \widetilde{\mathcal{E}}_{B,c})$
    $\widetilde{\mathcal{M}} \leftarrow (\mathcal{M} \setminus \mathcal{A}_s) \cup \{\hat{\mathcal{A}}\}$
    $\widetilde{T}, \widetilde{\mathcal{M}} = \text{HillClimbing}(\widetilde{\mathcal{M}}, \widetilde{\mathcal{E}}_{B,c})$
    **if** $\widetilde{T} > T$ **then**
        $T = \widetilde{T}, \quad \mathcal{M} \leftarrow \widetilde{\mathcal{M}}, \quad \mathcal{E}_{B,c} \leftarrow \widetilde{\mathcal{E}}_{B,c}$
**return** $T$, $\mathcal{M}$

The next step is the running of the *Rate Increase* algorithm (Algorithm 3). The rationale underlying it is the following: some cells, hence areas, are hampered by low-bit-rate users (likely those near the cell edge). These underper-forming users can be removed from the set of those served via broadcast in order to boost the bit rate of the synchronization area. Recall, indeed, that the bit rate of the broadcasting service of a specific item must match the lowest bit rate *among broadcast users* in the MBSFN area. In order to serve the target fraction of broadcast demand, users that are not served through eMBMS should receive the desired content via unicast. Thus, whether it is convenient to actually remove slow users or not has to be evaluated.

In the algorithm, $\mathcal{S}$ is the list of user rates in increasing order. For each rate value, the function *FindRateUsers* identifies the users requiring such data rate ($\mathcal{E}_s$) and returns their identity along with the areas ($\mathcal{A}_s$) to which they belong (line 2). These users are no longer taken into account and the Cell Aggregation algorithm is run again on the selected areas because the condition on interested users in Algorithm 1, line 5, may no longer hold. The areas may thus become smaller, or they may even need to be split in smaller portions. In the latter case, the selected areas are removed from the activation set $\mathcal{M}$ and replaced by such portions (line 5). The Hill Climbing algorithm is then run again on the new $\mathcal{M}$ and with the new set of users. If the throughput it returns is higher than before, the new settings are consolidated.

The final step is represented by the *Area Fusion* algorithm. Intuitively, its purpose is to merge fully- or partially-overlapping single-content areas, thus forming multiple-content areas, until constraint *(i)* on the maximum number of areas is met. Due to its simplicity, we briefly sketch it below, without resorting to pseudocode.

First, the algorithm merges fully-overlapping MBSFNs returned by Algorithm 3 so as to create multi-content areas. It then checks if MBSFNs exceed the maximum number dictated by constraint *(i)* in Sec. 3.2. If so, it proceeds to identify partially-overlapping area pairs, each broadcasting different content, and sorts them by increasing number of differing cells. Area Fusion merges the pair that both meets the capacity constraints *(iii)*-*(iv)* and provides the best throughput performance. This step is repeated until either constraint *(i)* is met or the throughput is degraded. Indeed, the trouble with partially-overlapping areas is that content is broadcast in areas where potentially no-one is interested in it. If this is the case, the algorithm stops and the first Max_MBSFN areas are kept. By exploiting similarity in spatial coverage, the Area Fusion algorithm thus manages to define areas that broadcast multiple, highly-requested content, satisfying all system constraints.

### *4.2. Complexity of the SCF procedure*

Decisions on MBSFN area formation have to be taken with a medium-to-low frequency, e.g., once every few hours. This means that time complexity is not as critical as in other applications (e.g., scheduling) where decisions are taken in real time. In spite of this, our algorithms have a very good level of complexity, which ensures their scalability and makes them suitable for usage in large-scale, real-world scenarios.

**Cell Aggregation.** In Algorithm 1, for every content, we make one decision for each cluster of cells whose size increases at each iteration. The number of content items is $|\mathcal{I}|$; the number of cell clusters that are processed is of the order of the number of annuli in the synchronization area with thickness equal to one cell, i.e., $\sqrt{|\mathcal{C}|}$. The total complexity is therefore $O(|\mathcal{I}|\sqrt{|\mathcal{C}|})$.

**Hill Climbing.** Selecting the areas to activate is potentially the most critical part of our solution. However, the Hill Climbing approach we adopt in Algorithm 2 exhibits a remarkably low complexity, namely, proportional to the square of the number $|\mathcal{A}|$ of areas to check. The worst-case complexity is then $O(|\mathcal{A}|^2)$, with typically $|\mathcal{A}| \ll |\mathcal{I}||\mathcal{C}|$. The price we pay for such a low complexity is a somehow loose optimality bound at $1 - \frac{1}{e}$ [18]; however, hill-climbing heuristics are commonplace when dealing with large-scale, complex problems, and are routinely found to work remarkably closer to the optimum than such a bound.

**Rate Increase and Area Fusion.** The Rate Increase algorithm runs the Hill Climbing procedure a number of times equal to the number of data rate values, which can be discretized and thus are in a constant finite number. It follows that the complexity of the two algorithms coincide. As for the Area Fusion, the highest complexity is due to the comparison between areas, which is $O(|\mathcal{M}|^2)$, with $|\mathcal{M}| \leq |\mathcal{A}|$.

In conclusion, the complexity of the SCF procedure is driven by that of the Hill Climbing algorithm, which is adequate to address scenarios covering large geographical regions.

## 5. The Multiple-Content Fusion procedure

The single-content fusion procedure described in Sec. 4 takes into account one content at a time. To understand the impact of such a feature on the decisions made by the algorithms and on the global performance, we benchmark the SCF procedure against a *multiple-content fusion* (MCF) procedure, presented in this section, which jointly accounts for multiple contents. The MCF procedure operates as follows.

(1) Upon startup, every cell is an individual area broadcasting the content items for which there are at least $\tau$ interested users in the cell, capacity constraints *(iii)*–*(iv)* are met and the system throughput is maximized.

(2) Then, for each area $m$, the neighboring area with the highest value of interest similarity is chosen as a candidate area for aggregation. The similarity metric is computed as the Euclidean distance between the area interests over the multidimensional space of the broadcast content items. If the aggregation meets constraints *(ii)*–*(iv)* and there exists a subset of common content items that leads to a throughput increase (as computed by the Hill Climbing algorithm), then the two areas are combined and the best content subset is selected. Otherwise, $m$ is no longer a candidate for aggregation. Note that such area processing may actually yield up to three areas: one broadcasting the selected content subset, and two the remaining (if any) content items of each member of the pair. All areas obtained as a result of this step are inserted in the pool of areas to be processed.

(3) When all areas have been examined, the Rate Increase algorithm is run, following the same principles expounded for the SCF scheme. If the number of formed areas exceeds the maximum (constraint *(i)*), only the most performing Max_MBSFN areas are activated. Indeed, the MCF algorithm already yields merged multi-content areas, thus further fusing the areas is not necessary (recall that all advantageous aggregations have been performed at step (2)).

As evident from the above description, while SCF tends to form large areas, each broadcasting few items, MCF adopts the opposite approach: it forms small areas, each broadcasting several content items. At last, we stress that the complexity of MCF is significantly higher than that of SCF, as now the hill-climbing algorithm has to decide, for each content $i \in \mathcal{I}$, whether to broadcast it over the area that it has processed or not. It turns out that the number of elements, on which the algorithm operates, is at least of the order of $|\mathcal{I}||\mathcal{C}|$ instead of $|\mathcal{A}|$. This results in a complexity that is at least $O(|\mathcal{I}|^2|\mathcal{C}|^2)$.

## 6. Numerical results

We evaluate the performance of our approach using an ad hoc simulator in Python, that:

- takes into account the reference scenario typically adopted in the literature, and described in Sec. 3.1;
- implements the SCF and MCF procedures, as described in Sec. 4 and Sec. 5, respectively;
- computes the throughput metrics described in Sec. 3.3 and Sec. 3.4, resulting from the decisions made by the SCF and MCF procedures.

In particular, we target the 57-cell scenario adopted by 3GPP for LTE network evaluation [19], as well as a larger-scale scenario with 597 cells. The synchronization area extends over about 4 km$^2$ in the small scenario, and 43 km$^2$ in the large one; the distance between any two eNBs is 500 m. All cells belong to the same synchronization area, and users are uniformly distributed therein, with an average density of 60 users per cell. In line with [20, 19], for eNBs we assume a transmit power of 43 dBm, an antenna height of 25 m and an antenna gain of 14 dBi. For the UE, we set the antenna height to 1.5 m and the antenna gain to 0 dBi. All nodes operate over a 10-MHz band at 2.6 GHz, thus we have $R = 500$ available RBs per frame. A $2 \times 2$ MIMO is considered. Signal propagation over BS-UE links is modeled according to ITU specifications for the Urban Macro environment [20], while the SINR is mapped onto per-RB throughput values using the experimental measurements in [21]. User positions and link propagation conditions are considered constant since they represent the average system behavior between two consecutive area formation procedures.

Each user is interested in one content item only. To represent location-based content, we divide the synchronization area into zones, in each of which users select the content they are interested in out of a set of 16 items. This value is motivated by the fact that the counting procedure allows interest collection for no more than 16 items at a time [2]. In the small-scale scenario, the number of zones is set to 4, while in the large-scale one it is a varying parameter. Also, the content set varies from a zone to another in such a way that, unless otherwise specified, a zone has a total of 12 (randomly selected) items in common with its neighboring zones. Within a single zone, user interest in the content set is either uniformly distributed or follows a truncated negative exponential distribution with parameter[2] 3.5. The latter value implies that there are four items with a probability of at least 0.1 to be requested by the generic user. The content service rate is a varying parameter. Finally, recall that the target fraction of broadcast demand that should be served (via either eMBMS or unicast) is set to 1, while the threshold $\tau$ on the number of interested users in Algorithm 1 is set to 2.

All results have been obtained with a 95% confidence interval. While showing the performance, we will present, among others, the following metrics.

**Broadcast/unicast serving ratio.** It is the ratio of the broadcast demand served via either eMBMS or unicast delivery, to the broadcast demand served through unicast only. This ratio is equal to 1 only for low traffic load (where it makes little sense to use broadcasting in the first place), while it exceeds 1 as the traffic load grows and an efficient allocation of radio resources becomes more and more important. In other words, the efficiency of eMBMS in accommodating the broadcast demand w.r.t. unicast LTE networks is represented by how much the serving ratio exceeds 1.

[2] In the case of exponential distribution, the 12 items a zone has in common with its neighbors are the least popular ones.

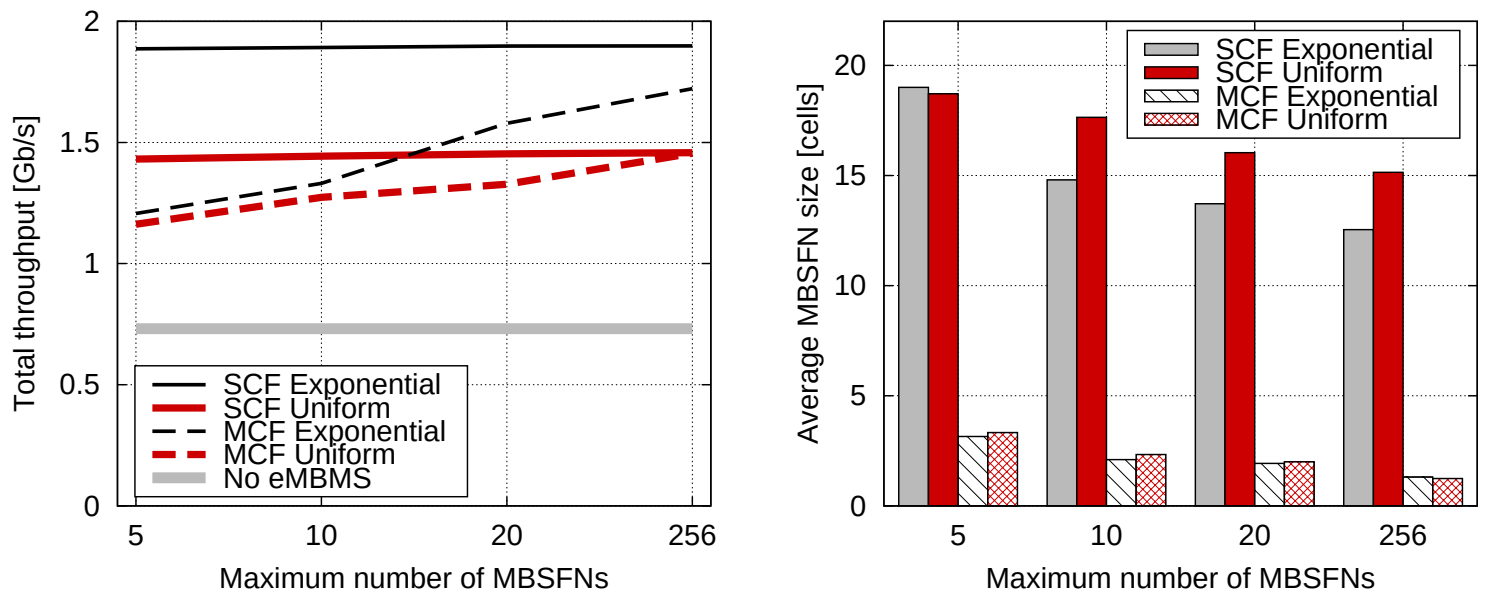

Figure 3: SCF vs. MCF in the small-scale scenario as Max_MBSFN varies. Total throughput (left) and average MBSFN size (right), for 4 interest zones and a service rate of the broadcast content equal to 500 kb/s.

**RB gain**. It is the ratio of the number of RBs used to serve broadcast demand when eMBMS is not active (i.e., only unicast service is allowed), to that used when eMBMS is active (i.e., both broadcast and unicast services are enabled). For a fixed value of broadcast/unicast serving ratio, the higher the RB gain, the more evident the benefit of eMBMS.

**Fraction of used RBs.** It is the fraction of total RBs per frame that are used for a given data transfer, i.e., broadcast demand via broadcast delivery (B(B)), broadcast demand via unicast delivery (B(U)), and unicast demand (U(U)).

### *6.1. Static scenario*

We start by comparing the SCF to the MCF scheme in a static scenario, where user interest does not change over time. We aim at confirming our intuition that assessing cell interest similarity on a content-by-content basis leads to better system configuration than acting by considering several content items at the same time. The comparison is made in the small-scale scenario as the complexity of the MCF scheme prevents us from dealing with larger instances in a reasonable amount of time. Here, the service rate of the broadcast items is assumed to be 500 kb/s for all of them. Due to the small scenario size, a sensible comparison between SCF and MCF is done by considering different values of Max_MBSFN. This parameter is typically set to 256, but considering we have only 57 cells in total, we scale this parameter down.

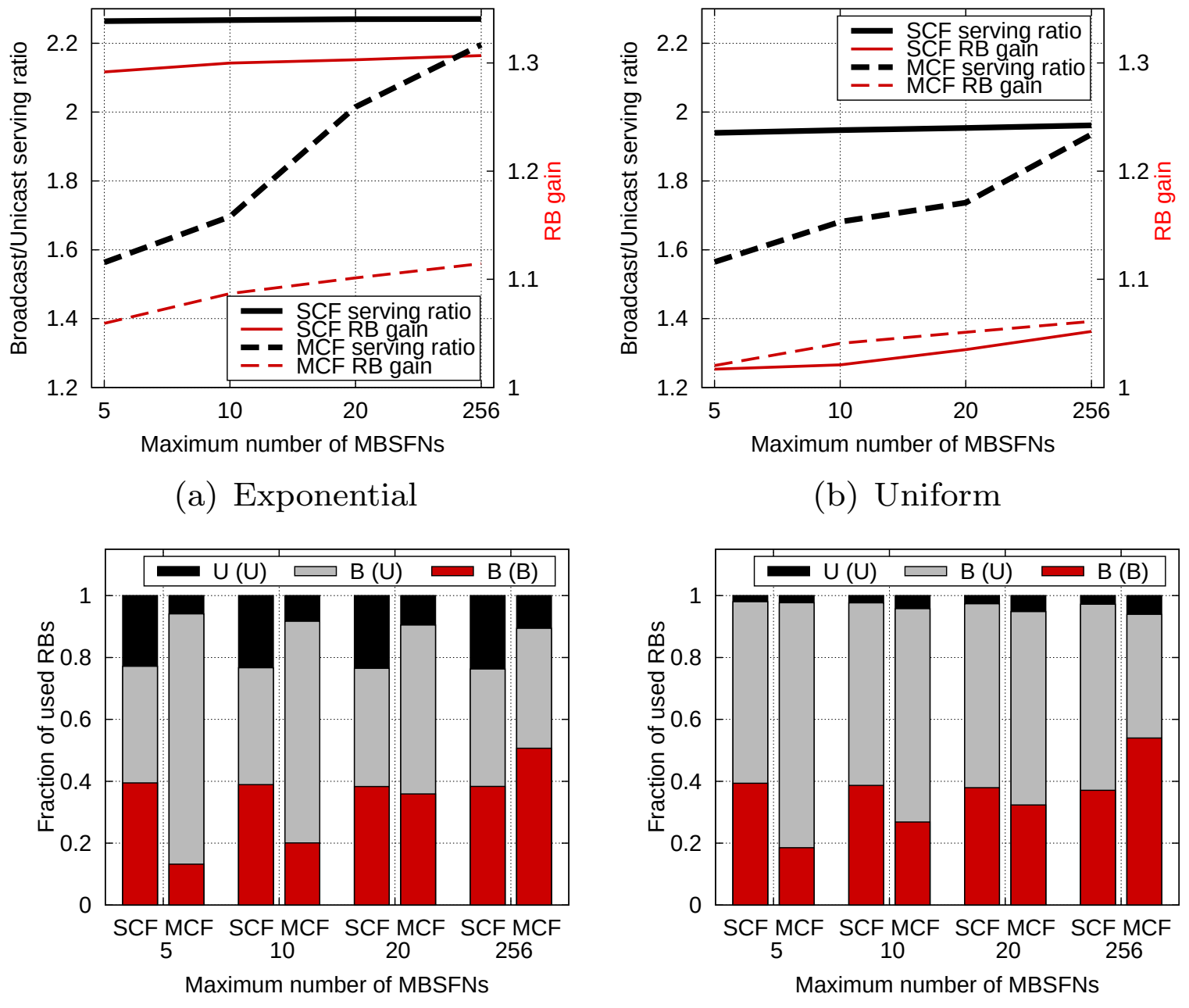

(a) Exponential (b) Uniform

(c) Exponential (d) Uniform

Figure 4: Comparison between SCF and MCF in the small-scale scenario vs. Max_MBSFN, with 4 interest zones and service rate of the broadcast content equal to 500 kb/s. Served users ratio and RB gain (a),(b), and fraction of used RBs (c),(d) for different distributions of the user interest.

Fig. 3 depicts the total throughput, $T$ in equation (4), over the synchronization area and the average MBSFN size when eMBMS is active and either SCF or MCF is applied as well as when only unicast is possible. Results are shown for both uniform and exponential interest distribution. As expected, using eMBMS leads to much better throughput than unicast traffic delivery only (left plot). Looking at the comparison between SCF and MCF, the SCF through-put remains almost constant, and always higher than MCF, as the maximum number of MBSFNs varies. Indeed, SCF can reach the maximum throughput by activating a small number of large areas (right plot) – recall that large areas can leverage the constructive interference among cells belonging to the same MBSFN.

MCF instead tends to activate many MBSFNs, each composed of very few cells (right plot) and broadcasting several items. The reasons for this behavior are as follows: (i) MCF selects the content items to be broadcasted in an area more efficiently as the number of aggregated cells is small, and (ii) small MBSFNs allow the broadcast transmission rate to be tailored to a limited set of users, thus allowing, on average, a higher rate than in the case of large areas. It follows that the MCF throughput gets close to that provided by SCF (left plot) only for the anomalous case where the areas formed by MCF are essentially all single-cell MBSFNs (right plot for Max_MBSFN equal to 256).

Finally, it is important to note that, when the user interest is exponentially distributed over all possible items, much better results are obtained than in the case of uniform distribution, and this is particularly evident for SCF. To wit, the more the user interest is concentrated, for each cell, on few items, the larger the broadcast demand that each area created by SCF can serve. Thus, limiting the number of popular items ensures that the allocation of few RBs will be enough to satisfy a large number of users.

Fig. 4 depicts the broadcast/unicast serving ratio and RB gain defined above, when interests are exponentially (a) and uniformly distributed (b) over content items. First we note that, again, both SCF and MCF provide much better performance than unicast delivery. Indeed, even when the RB gain is close to 1, the serving ratio of both schemes is significantly higher than that.

Focusing on the SCF vs. MCF comparison, in case of exponentially distributed interests, SCF remarkably outperforms MCF in terms of both metrics: not only can SCF serve more users but it also allocates radio resources more efficiently. The latter is due to the fact that SCF can create large areas and serve most of the broadcast demand through eMBMs, i.e., lots of users served by a small number of RBs. This is also evident from Fig. 4(c), which shows the fraction of used RBs and highlights that SCF leaves significantly more room for unicast demand.

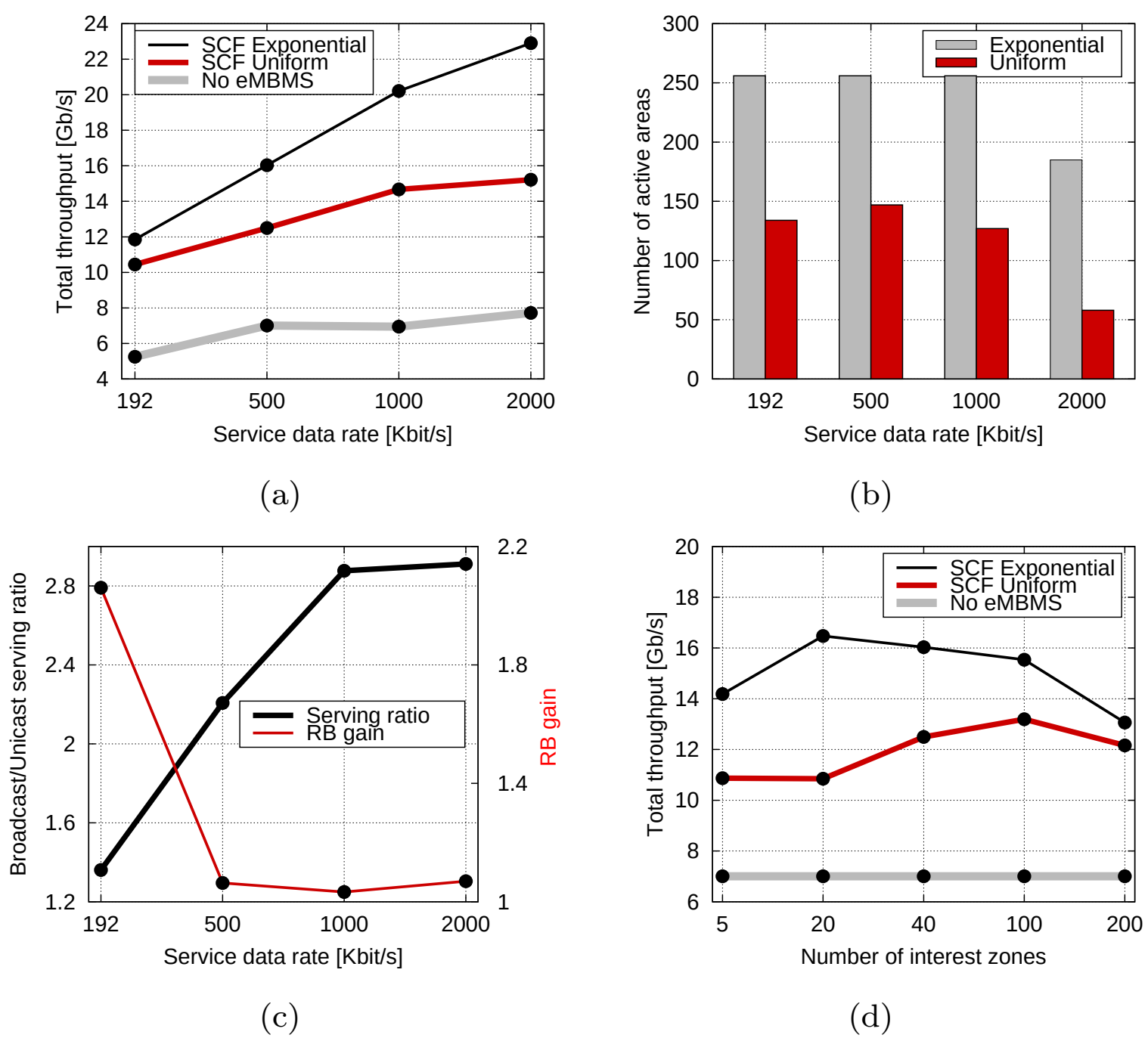

Figure 5: SCF performance in the large-scale scenario. The first 3 plots on the left refer to 40 interest zones and different service rates of the broadcast content: Total throughput (a), number of activated MBSFNs (b), served users ratio and RB gain for exponentially-distributed user interest (c). The rightmost plot (d) shows the throughput for a service rate of 500 kb/s and as the number of interest zones varies.

The gray area in the SCF bars in Fig. 4(c) corresponds to the RBs that are devoted to delivering broadcast content to users that are not part of any MBSFN and, to a lesser extent, to slow users that have been removed from the eMBMS by the Rate Increase algorithm. In MCF, for low-medium values of Max_MBSFN, the small areas that are created can serve just a limited number of users while most of them will receive content via unicast delivery, as shown in Fig. 4(c). It follows that MCF can set the broadcast data rate higher, thus requiring very few RBs for the eMBMS transmissions. However, it is important to remark that, while SCF can serve 100% of the broadcast demand, MCF fails to serve from 30% to 3% of such a demand (for a maximum number of MBSFNs ranging between 5 and 256), leaving out the slowest users.

In case of users interested in content items with equal probability (see Fig. 4(b)), a different behavior emerges: SCF serving ratio and RB gain are always, respectively, higher and lower than those of MCF. SCF initially forms large areas (via Cell Aggregation and Hill Climbing) but very few users per cell are interested in the same content. Also, for low values of Max_MBSFN, such areas cannot be broken into smaller portions by the Rate Increase algorithm (recall that when not enough users are left in the cells to justify content broadcasting therein, the cells themselves have to be removed from the MBSFN). Too large areas imply a higher probability of having slow users to serve, which translates into a lower broadcast transmission rate and a higher number of RBs that SCF has to allocate for content broadcasting (see Fig. 4(d)). On the contrary, as noted before, the small areas of MCF can broadcast at high rate. As the number of allowed MBSFNs grows, the Rate Increase algorithm in SCF tends to break large areas into smaller ones (see also the right plot of Fig. 3), leading to RB gain performance that is closer to that obtained with MCF. What is important to observe, however, is that the worse RB usage in SCF does not hurt unicast demand more than what MCF does. As for the number of unserved users, SCF greatly outperforms MCF: 10% vs. 30%–12%. In summary, MCF performance matches that of SCF only in the anomalous case where MCF forms single-cell MBSFNs (i.e., for Max_MBSFN equal to 256).

The rest of the results depict the performance of our SCF scheme in the large-scale scenario. First, we fix the number of interest zones to 40, Max_MBSFN to 256, and we test 4 different broadcast service rates (Figs. 5(a)–(c)). The chosen service rates represent typical bandwidth demands for mobile video services at various screen resolutions, assuming the recent H.264/AVC compression technology [22]. Clearly, the case where all broadcast items have a service rate of 2 Mb/s corresponds to a huge traffic load, for which we expect that the radio resources that can be allocated for eMBMS will be insufficient to serve the whole broadcast demand.

The total throughput in Fig. 5(a) again highlights the significant gain achieved w.r.t. the case where no eMBMS is used. Also, the larger the service rate, the higher the throughput as the gain due to broadcast delivery increases. Indeed, recall that the per-frame broadcast throughput is multiplied by the number of users served through eMBMS (i.e., factor $w_c^i$ in (1)), thus the higher throughput is evidence of such a multiplicative effect. Interestingly, the gain is more remarkable when users in a zone are mostly interested in the same subset of content items (i.e., for exponential distribution). This is because, as shown by Fig. 5(b), a high service rate implies that very few items can be broadcasted per cell, due to constraint *(iii)*. Thus, the number of overlapping MBSFNs that can be formed is significantly reduced. Consistently, the number of active areas in the right plot decreases for both exponentially and uniformly distributed user interest. In the exponential case, however, there are many more users interested in the content item that is broadcasted than in the uniform case (thus factor $w_c^i$ takes larger values in the former scenario). This leads to a higher broadcast gain in case of exponential distribution of user interest.

Fig. 5(c) shows the serving ratio and RB gain, as well as the fraction of used RBs when user interest is exponentially distributed over the set of broadcast items. (For brevity, the uniform case is omitted as it exhibits exactly the same qualitative behavior.) Observe that, for low broadcast traffic load, unicast delivery and eMBMS manages to support the broadcast demand almost equally, although the latter needs fewer RBs. As the broadcast service rate increases, saturation is soon reached by unicast (unserved users range from 25% at 192 kb/s to 90% at 2 Mb/s). SCF instead can successfully support the broadcast demand till a service rate of 500 kb/s and can still serve 40% of users in the extreme case where all items have a service rate of 2 Mb/s. Clearly, when the two curves intersect we have the best trade-off between serving ratio and RB gain.

In Fig. 5(d), we fix the service rate to 500 kb/s and vary the number of user interest zones. The SCF throughput, which is much higher than in the case without eMBMS, initially grows with the number of zones. This is due to zones becoming smaller, which yields an increase in the variety of requested broadcast items over the synchronization area. A larger number of relatively small MBSFNs are thus formed. As MBSFNs get smaller, the average broadcast data rate increases, the Rate Increase algorithm is more efficient, and the resulting throughput is higher. Upon a further increase in the number of zones, the performance dips. In this case, the number of MBSFNs grows significantly till it hits the maximum number of allowed areas. Consequently, not all MBSFNs that could be created can be actually activated. Furthermore, MBSFNs become too small and lose the benefit coming from constructive interference. Thus, the plot highlights an interesting trade-off on the MBSFN size: too large areas lead to low broadcast rates, while small areas do not permit the exploitation of constructive interference and their number may easily exceed the maximum.

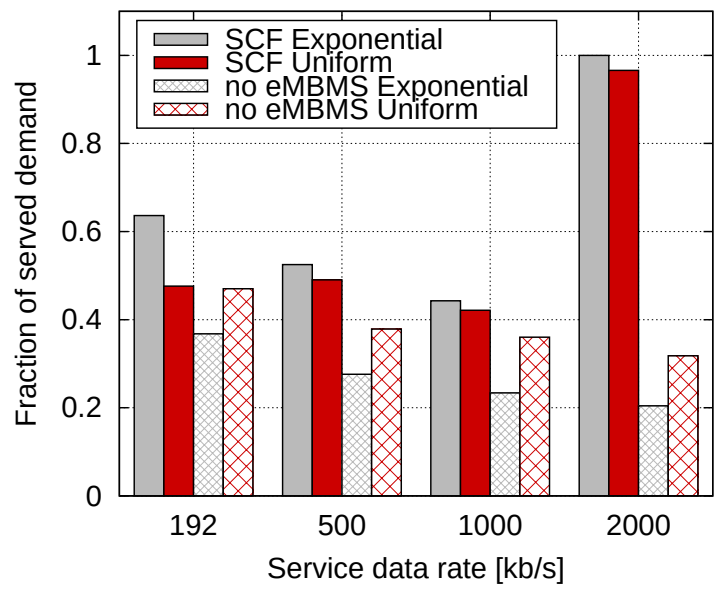

Figure 6: Fraction of served broadcast demand in the large-scale scenario when items with different service rates can be requested: SCF vs. unicast delivery for exponential and uniform user interest distribution.

Finally, Fig. 6 presents the performance in the case where the system is saturated and users can request content items with different service rates (the number of zones is set to 40 as before). In particular, we consider a 2-Mb/s item representing a live streaming event that may be of interest to all users in the synchronization area. Then, in each zone, users may be interested in the same (but different from zone to zone) 1-Mb/s item, or in one out of 14 small items whose service rate is randomly selected between 192 kb/s and 500 kb/s.

Interestingly, the plot shows that SCF can serve almost all users requesting the highly popular live streaming event and most of the other user requests.

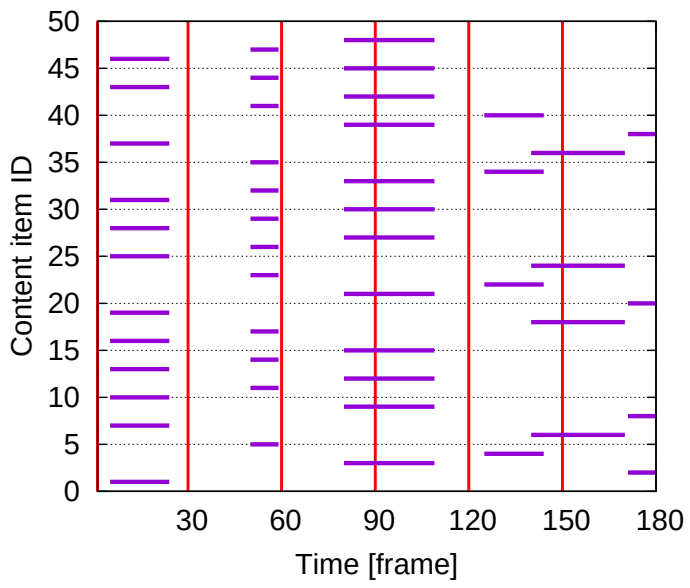

Figure 7: Activation timeline for different content items (IDs shown on the y axis).

With regard to the latter, slightly higher priority is given to small content items as SCF aims at maximizing the system throughput, which also depends on the number of users interested in a specific content. Indeed, for a fixed number of interested users per item, it is more convenient to broadcast the small ones first as more can be accommodated. Finally, while unicast delivery may be quite effective in serving small content, it is no match for SCF when large items should be served. This is also evident from unicast delivery achieving better performance in the case of uniform user interest distribution than in the exponential case. Indeed, under the uniform assumption, fewer users per cell select large content items than in the exponential case. As a result, fewer resources have to be spent for such items and more requests for small-sized content can be satisfied.

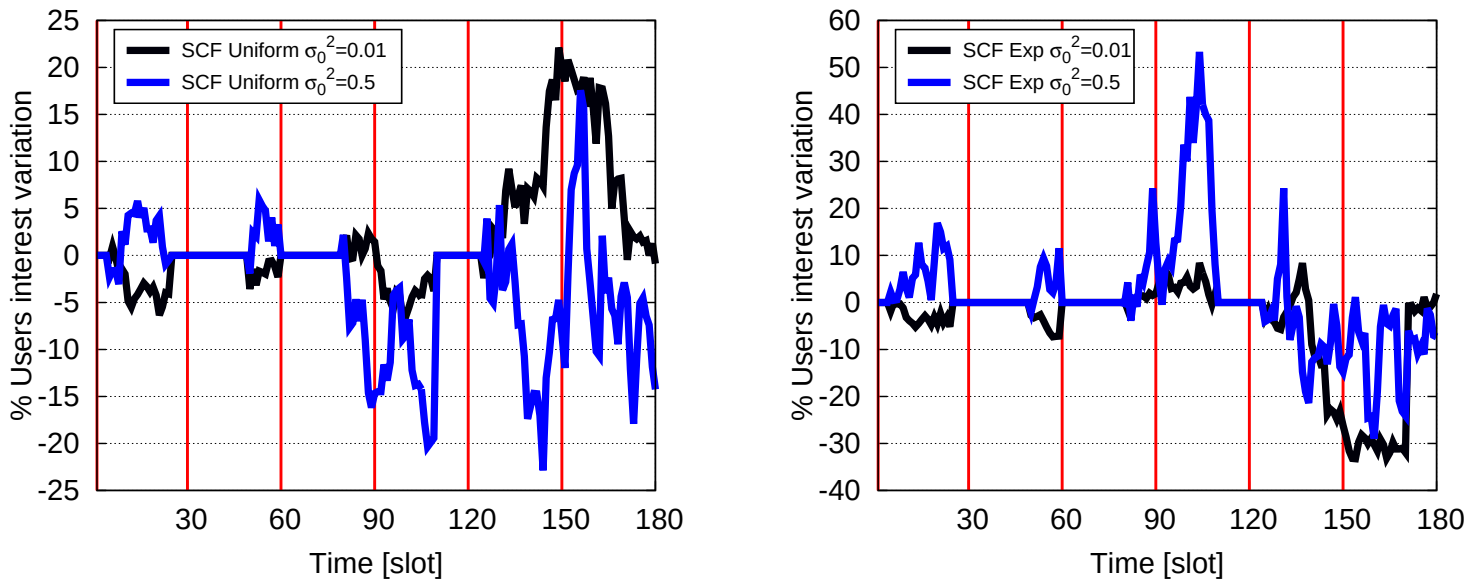

(a) Uniform

(b) Exponential

Figure 8: Example of the percentage variations in the number of users interested in a generic content item, as a function of time, for uniform (a) and exponentially-distributed (b) user interest and different $\sigma_0^2$.

*6.2. Dynamic scenario*

We evaluate the performance of a dynamic scenario in the 57-cell (small scale) setting that was previously described. As shown in Fig. 7, we assume that content items are temporally distributed over a period of 180 time frames, each lasting one minute. The period $P$ between two consecutive SCF algorithm runs is assumed to be 30 time frames, and is shown by red vertical lines in the figure. Content items have different random duration, uniformly selected between 10 and 30 time frames and can start at any time.

In the real world, the exact number of users interested in a content item is likely unknown, but it can be more and more accurately estimated as the content start time approaches. For instance, an operator may perform a survey procedure as foreseen by the eMBMS standard, or it may derive a rough figure looking at the number of users who have logged into a streaming service minutes ahead of the start of a sport event. Recommendation systems also influence the number of users interested in a content item, in a predictable manner. Hence, when the SCF algorithm is run (e.g., periodically, as we assume here), the estimate may be closer to the actual number depending on how long it is before the content start time. To account for such uncertainty, in our simulations we shall compute $\tilde{w}_c^i$ in eq. (5) by taking the actual number of users interested in content item $i$ and add a noise $\nu_{i,s}$ represented by an i.i.d. Gaussian-distributed variable with zero mean and variance $\sigma_s^2$. The variance is computed as $\sigma_s^2 = \sigma_0^2$ $t_s$, where $t_s$ is the time difference between the content start and the scheduling instant preceding it.

Furthermore, during the content broadcasting, we model the instantaneous interest variation due to users starting and then quitting the content fruition by an additional noise $\nu_{i,d}$. This is represented by an i.i.d. Gaussian-distributed variable with zero mean and variance $\sigma_d^2 = \sigma_{0,i}^2 t_d$, where $t_d$ is the residual duration of the content item (from the moment the decision is made). We thus expect that short content items (or items that are almost at the end of their run) exhibit a low variation in users interest while long-lasting items have a higher user churning rate. To avoid having to deal with negative numbers, $\nu_{i,s}$ is truncated in the range $[-w_c^i, w_c^i]$, while $\nu_{i,d}$ is truncated in the range $[-0.3w_c^i, 0.3w_c^i]$. An example of the temporal percentage variations in the number of users interested in a generic content item is shown in Fig. 8.

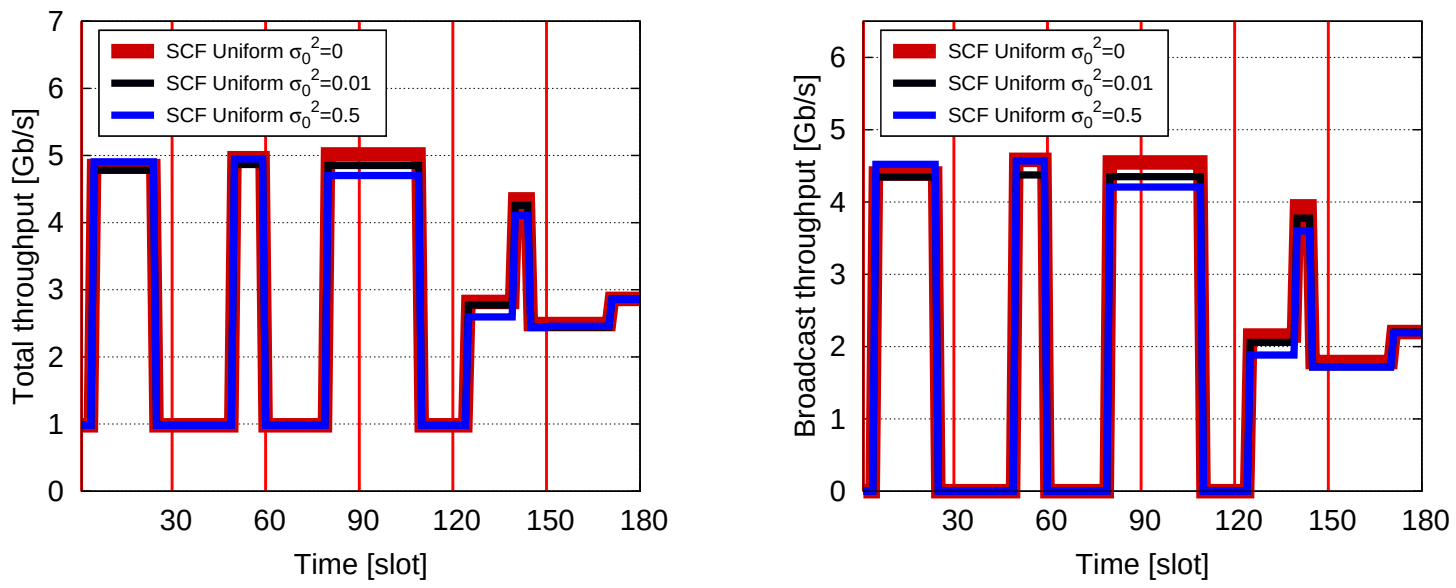

Figure 9: Total throughput (top) and broadcast throughput (bottom) for uniform content distribution and different $\sigma_0^2$; case of 2 Mb/s service rate.

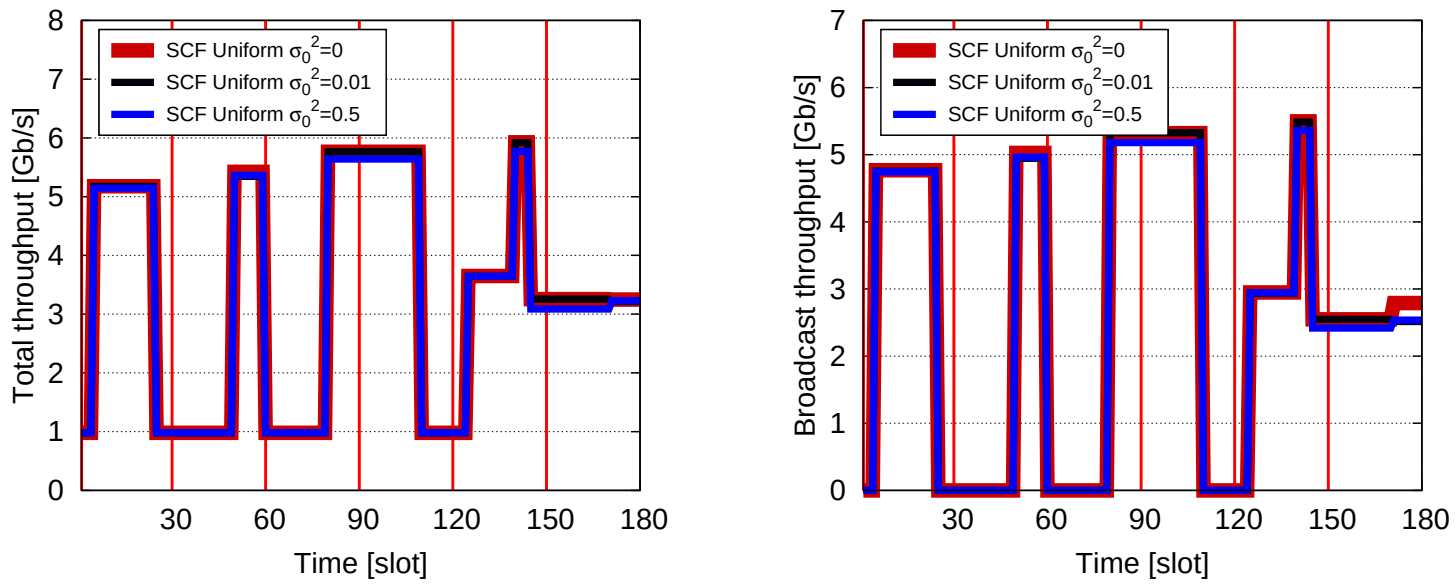

Figure 10: Total throughput (top) and broadcast throughput (bottom) for exponential content distribution and different $\sigma_0^2$; case of 2 Mb/s service rate.

We tested several configurations in terms of service rate of content items, ranging from 500 kb/s to 2 Mb/s, as in the static scenario. Each yielded similar results, therefore in the following we just discuss the case of 2 Mb/s rate. The overall throughput and the broadcast throughput over time are shown in Figs. 9 and 10, for different values of $\sigma_{0,i}^2 = \sigma_0^2$, assumed to be identical for all content items ($\sigma_0^2 = 0$ corresponds to the case of perfect prediction). The achieved throughput matches quite accurately the interest variations and shows that SCF can efficiently schedule broadcast flows even in dynamic settings. Although there is a slight throughput loss for higher values of $\sigma_0^2$, hence of uncertainty, the fact that SCF operates on *groups* of users, rather than on single users, makes it more unlikely that its decisions are offset by the uncertainty in user interests. Similar observations are also in order for both the static and the dynamic case: in both exponential interest translates into higher throughput, due to the reasons discussed in the previous subsection.

## 7. Conclusions

One of the most promising ways to efficiently deliver popular content to a large number of users without overloading the network is embodied by LTE broadcasting. In this paper, we addressed the crucial problem of efficient broadcast area formation based on existing cell topology and the interests of users scattered across cells. Specifically, we proposed an area formation procedure, called SCF, that operates by (i) aggregating cells into several potential single-content MBSFNs based on user interests; (ii) through a hill-climbing approach, selecting which single-content areas should be activated and which users MBSFNs should serve to maximize the overall throughput; (iii) creating multi-content areas by merging those that significantly overlap in space until system constraints are satisfied.

Performance evaluation in standard scenarios proved the superiority in system configuration, hence throughput, of our solution with respect to acting by considering several content items at the same time. We also discussed and analyzed the use of SCF for area formation in dynamic scenarios with changing interests, highlighting its robustness with respect to uncertainty in user interest variations.

## Acknowledgment

The research leading to these results was supported through the 5G-Crosshaul project (grant agreement no. 671598) and the I-REACT project (grant agreement no. 700256).